\documentclass[]{spie}  %>>> use for US letter paper
\usepackage[dvips]{graphicx}
\usepackage{amsmath,amstext,amsxtra,amsopn,amsgen,amsbsy}
\newcommand{\boomn}{{\sc Boomerang}}
\newcommand{\boom}{{\sc Boomerang} }
\newcommand{\apjl}{{\bf ApJL}}

\newcommand{\planckn}{{\it Planck}}
\newcommand{\planck}{{\it Planck }}
\newcommand{\planckhfi}{{\it Planck} HFI }
\newcommand{\planckhfin}{{\it Planck} HFI}
\newcommand{\sini}{$\mathrm{Si}_3\mathrm{N}_4$ }

\title{A Polarization Sensitive Bolometric Detector for Observations of the Cosmic Microwave Background
}

\author{W.C.~Jones\supit{a},~R.S.~Bhatia\supit{a},~J.J.~Bock\supit{a,b},~A.E.~Lange\supit{a}
\skiplinehalf
\supit{a}California Institute of Technology, Pasadena, CA, USA \\
\supit{b}Jet Propulsion Laboratory, Pasadena, CA, USA}
\authorinfo{Corresponding Author:\\
William Jones, 1200 E. California Blvd., MC 59-33, Pasadena CA
91125. email: wcj@astro.caltech.edu, tel: 626.395.2020
\\
\\
\copyright ~2002 Society of Photo-Optical Instrumentation
Engineers. This paper will be published in the SPIE {\emph
Proceedings} and is made available as an electronic preprint with
permission of SPIE.  One print or electronic copy may be made for
personal use only.  Systematic or multiple reproduction,
distribution to multiple locations via electronic or other means,
duplication of any material in this paper for a fee or for
commercial purposes, or modification of the content of the paper
are prohibited.
\\}
\begin{document}
\maketitle

%%%%%%%%%%%%%%%%%%%%%%%%%%%%%%%%%%%%%%%%%%%%%%%%%%%%%%%%%%%%%
\begin{abstract}

We have developed a bolometric detector that is intrinsically
sensitive to linear polarization which is optimized for making
measurements of the polarization of the cosmic microwave
background radiation. The receiver consists of a pair of
co-located silicon nitride micromesh absorbers which couple
anisotropically to linearly polarized radiation through a
corrugated waveguide structure. This system allows simultaneous
background limited measurements of the Stokes $I$ and $Q$
parameters over $\sim 30$\% bandwidths at frequencies from $\sim
60$ to 600 GHz. Since both linear polarizations traverse identical
optical paths from the sky to the point of detection, the
susceptibility to systematic effects is minimized. The amount of
uncorrelated noise between the two polarization senses is limited
to the quantum limit of thermal and photon shot noise, while
drifts in the relative responsivity to orthogonal polarizations
are limited to the effect of non-uniformity in the thin film
deposition of the leads and the intrinsic thermistor properties.
Devices using NTD Ge thermistors have achieved NEPs of $2 \cdot
10^{-17} ~ \mathrm{W}/\sqrt{\mathrm{Hz}}$ with a $1/f$ knee below
100 mHz at a base temperature of 270 mK. Numerical modelling of
the structures has been used to optimize the bolometer geometry
and coupling to optics. Comparisons of numerical results and
experimental data are made. A description of how the quantities
measured by the device can be interpreted in terms of the Stokes
parameters is presented. The receiver developed for the \boom and
\planckhfi focal planes is presented in detail.
\end{abstract}


%%%%%%%%%%%%%%%%%%%%%%%%%%%%%%%%%%%%%%%%%%%%%%%%%%%%%%%%%%%%%

\section{INTRODUCTION}
\label{sect:intro}  % \label{} allows reference to this section

Observational cosmologists have yet to detect polarization in the
cosmic microwave background radiation (CMB), and upper limits are
still well above the level expected due to Thompson scattering of
quadrupole anisotropies in the background radiation during the
epoch of recombination.\cite{staggs99,odell02}  The small
amplitude of this polarized signal, peaking at perhaps $5 \mu$K at
$\sim 10'$ scales, demands not only extremely high raw
sensitivity, but also exquisite control of systematics.

\noindent Over the last twenty years, nearly all published efforts
to detect polarization in the CMB have used coherent
receivers.\footnote{To the author\rq s knowledge, the pioneering
efforts of Caderni, et al., are the only published CMB
polarization limits set by a bolometric system.\cite{caderni}}
Heterodyne, quasi total-power, and correlation receivers with
front-end RF low noise amplifier blocks based on HEMTs are mature
technologies at millimeter wavelengths. The fundamental design
principles of these of receivers are well established and have
been used to construct polarized receivers at radio to mm-wave
frequencies for many years.\cite{gaier,spiga02} Although cryogenic
bolometric receivers achieve far higher instantaneous
sensitivities over wider bandwidths than their coherent analogs,
the intrinsic polarization sensitivity of coherent systems has
made them the choice of the first generation of CMB polarization
experiments.

\noindent In this paper we describe a new bolometric system which
combines the sensitivity, bandwidth, and stability of a cryogenic
bolometer with the intrinsic polarization capability traditionally
associated with coherent systems. In addition, the design obviates
the need for orthogonal mode transducers (OMTs), hybrid tee
networks, waveguide plumbing, or quasi-optical beam splitters
whose size and weight make fabrication of large format arrays
impractical. Finally, unlike OMTs or other waveguide devices,
these systems can be relatively easily scaled to $\sim 600$ GHz,
limited at high frequencies only by the ability to reliably
manufacture sufficiently small single-moded corrugated structures.
Polarization sensitive bolometers (PSBs) are fabricated using the
proven photolithographic techniques used to produce \lq spider
web\rq ~ bolometers, and enjoy the same benefits of reduced heat
capacity, negligible cross section to cosmic rays, and structural
rigidity.\cite{yun}

\noindent Polarization sensitivity is achieved by controlling the
vector surface current distribution on the absorber, and thus the
efficiency of the ohmic dissipation of incident Poynting flux.
This approach requires that the optics, filtering, and coupling
structure preserve the sense of polarization of the incident
radiation with high fidelity. A multi stage corrugated feed
structure and coupling cavity has been designed which achieves
polarization sensitivity over a 33\% bandwidth. A next generation
of sub-orbital, ground based, and orbital bolometric CMB
polarization experiments, including \boomn, BICEP, QUEST, and the
\planckhfi are basing their receiver designs around the PSB
concept.

\begin{figure}[tbp]
\centering \rotatebox{0}{\scalebox{1}{\includegraphics{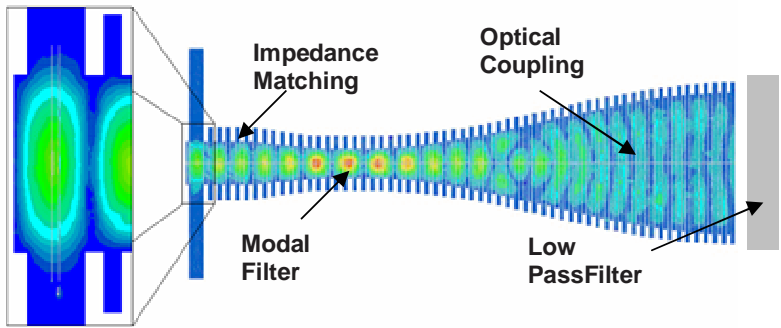}}}
\caption{An instantaneous image of the field distribution in
\boomn\rq s corrugated coupling feed. The radiation is incident
from the right where low-pass filters and, in some applications,
additional optical elements are located. The two bolometers are
symmetrically spaced at $\lambda_\mathrm{g} / 4 ~ +(-) ~ 30\mu$m
from the backshort in order to maximize coupling efficiency.
Similar feed structures have been designed for \planckn, QUEST and
BICEP\cite{kiwon,bicep} at 100, 150, 217, and 350 GHz.}
\label{fig:feedgeom}
\end{figure}

\section{General Design}

The PSB design has been driven by the desire to minimize
systematic contributions to the polarized signal.  Both senses of
linearly polarized radiation propagate through a single optical
path and filter stack prior to detection, thereby assuring both
detectors have identical spectral pass bands and quantum
efficiencies.  This results in nearly identical background loading
and closely matched responsivities between the two detectors. Two
orthogonal free standing lossy grids, which are separated by $\sim
60 \mu$m and both thermally and electrically isolated, are
impedance matched to terminate a corrugated waveguide structure.
The physical proximity of the two detectors assures that both
devices operate in identical RF and thermal environments.  A
printed circuit board attached to the module accommodates load
resistors and RF filtering on the leads entering the cavity.  The
post detection electronics consist of a highly stable AC readout
with a system $1/f$ knee below 100 mHz.  Unlike coherent systems,
this low frequency stability is attained without phase switching
the RF signal.

\subsection{Optics}
\label{sec:optics}

\noindent We have designed the optical elements, including the
feed antenna and detector assembly, to preserve sky polarization
and minimize instrumental polarization of unpolarized light.  To
this end, the detector has been designed as an integral part of
the optical feed structure.

\noindent Corrugated feeds couple radiation from the telescope to
the detector assembly.  Corrugated horns are the favored feed
element for high performance polarized reflector systems due to
their superior beam symmetry, large bandwidth, and low sidelobe
levels. In addition, cylindrical corrugated feeds and waveguide
preserve the orientation of polarized fields with higher fidelity
than do their smooth walled counterparts.

\begin{figure}[tbp]
\centering \rotatebox{0}{\scalebox{1}{\includegraphics{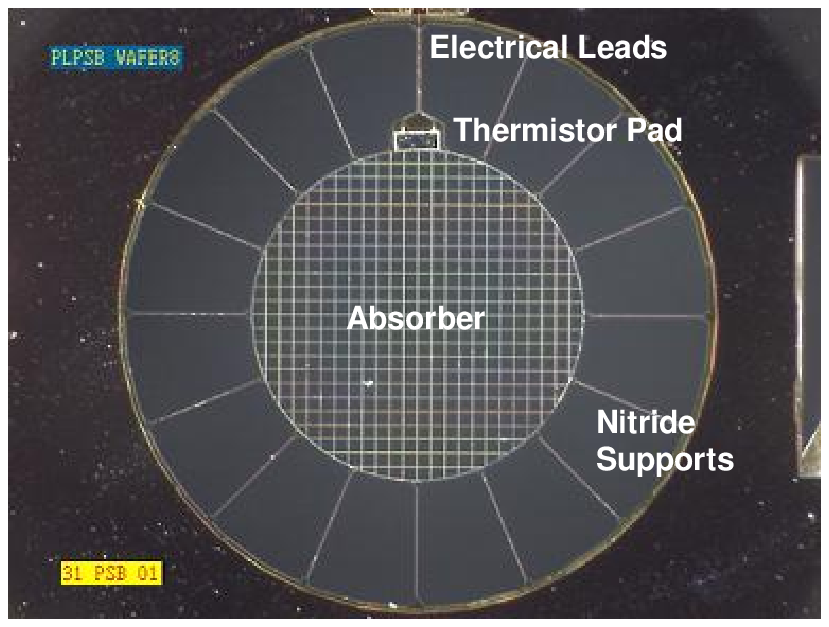}}}
\caption{A photograph of a \boom PSB absorber.  The diameter of
the grid is 2.6mm, while the absorber leg spacing, $g$, is 108
$\mu$m. Each leg is $3 \mu$m wide.  This device is sensitive to
incident radiation polarized in the vertical direction due to the
metallization of the \sini mesh in that direction. The horizontal
\sini beams evident in the photo are un-metallized, and provide
structural support for the device. The thermal conductivity
between the absorber and the heat sink is dominated by the
metallic leads running to the thermistor chip.}
\label{fig:absgeom}
\end{figure}

\noindent The coupling structure, which is cooled to 300 mK,
consists of a profiled corrugated horn, a modal filter, and an
impedance matching section which allows efficient coupling to the
polarization sensitive bolometer (see Figure \ref{fig:feedgeom}).
In addition to a reduction in the physical length of the
structure, the profiled horn provides a nearly uniform phase front
which couples well to the other filters and optical elements in
the system. The modal filter isolates the detectors from any
unwanted higher order modes that may be excited in the thermal
break. Equivalently, this filter completely separates the design
of the bolometer cavity from that of the feed which couples to the
optics. The impedance matching section (the re-expansion at the
left side of Figure \ref{fig:feedgeom}) produces a uniform vector
field distribution with a well defined guide
wavelength\footnote{The guide wavelength, $\lambda_\mathrm{g}$, is
 typically 20\% larger than free space, and
 $\mathrm{d}\log(\lambda_\mathrm{g})/\mathrm{d}\log(\nu)$ remains small over
the entire range of operation.} and characteristic impedance over
a large ($\sim 33\%$) bandwidth.

\subsection{Detector assembly}

\noindent The detector assembly is a corrugated waveguide
terminated with impedance-matched loads which have a weak thermal
link to the temperature bath. An electric field drives currents
through the absorber surfaces which result in ohmic power
dissipation. This power is detected as a temperature rise measured
by means of matched Neutron Transmutation Doped (NTD) Germanium
thermistors.\cite{beeman} The bolometers each couple to a single
(mutually orthogonal) linear polarization by properly matching
each absorber geometry to the vector field of the coupling
structure.  The coupling structure has been tailored to ensure
that the field distribution at the location of the bolometer
resulting from a polarized source is highly linear. The detector
assembly forms an RF tight Faraday cage around the bolometers, and
both the signal and bias lines are buffered by onboard RF filters.

\noindent Because the absorber geometry influences the field
distribution within the coupling structure, a treatment of the
bolometer cavity as a black body is in general {\em not}\, valid.
An important consequence of this fact is that any attempt to model
an analogous multi-moded optical system will have to carefully
consider interference terms between modes when calculating
coupling efficiencies or simply trying to predict radiation
patterns. The amplitude and phase of any higher order modes
capable of propagating to the bolometer depend on the details of
both the excitation and structure. Therefore, any numerical
calculation would be susceptible to a large number of
uncertainties associated with the appropriate boundary conditions
at the bolometer, while an accurate analytic solution would be
exceedingly difficult.  For this reason, it may prove difficult to
extend the general single mode PSB design to a multi-moded
application without sacrificing cross polar performance.

\begin{figure}[tbp]
\centering \scalebox{1.4}{\includegraphics{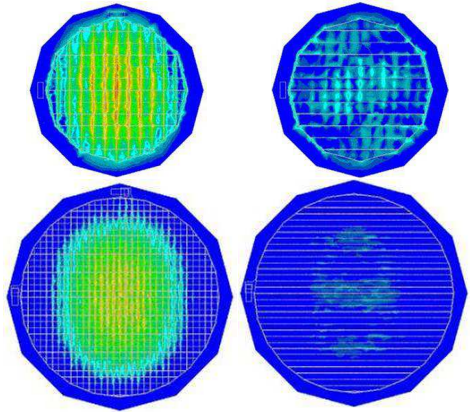}}

%\rotatebox{90}{\scalebox{.55}{\includegraphics{fullfeed2.ps}}}
%\scalebox{1}{\rotatebox{0}{\includegraphics{fig3a.eps}}\hspace{16mm}
%\hspace{16mm}\rotatebox{0}{\includegraphics{fig3b.eps}}}
%
%\hspace{2mm}\scalebox{1}{\rotatebox{0}{\includegraphics{fig3c.eps}}
%\rotatebox{0}{\includegraphics{fig3d.eps}}}

\caption{Plots of the Poynting flux through the surface of the
bolometer for the prototype PSBs (top) and the 143 GHz PSBs
designed for the \planckhfi (bottom). For each type of detector,
the leftmost plot corresponds to the co-polar device, while the
rightmost plot is the cross-polar device. The source polarization
is vertical. The intensity and physical scale is the same for all
four plots. The superior cross polar performance of the \planck
device is visually evident in the reduced cross-polar Poynting
flux. Integration of the component of the Poynting vector normal
to a surface completely inclosing each of the bolometers provides
a numerical estimate of the polarization efficiency,
$\varpi=1-\epsilon$. For the devices shown above the results are
shown in Figure \ref{fig:xpol_bands}, and correspond to cross
polar leakages, $\epsilon$, of about 5\% (prototype) and 1.5\%
(\planckn) across the band. Color, animated versions of the above
figures, and others, are available from the author upon request.}
\label{fig:pflux}
\end{figure}

\subsection{Absorber}

\noindent  Each bolometer consists of a linear absorber geometry
designed to couple independently to a single linear polarization.
PSBs are made using fabrication techniques originally developed
for the successful \lq spider-web\rq\, bolometers developed at
JPL.\cite{yun,jjb95} They consist of a free standing mesh of low
stress \sini with a $\sim 120$ \AA ~ thick layer of gold deposited
on $\sim 20$ \AA ~ of titanium as a thin film absorber. The
thermistor is located at the edge of the absorbing mesh, well into
the groove of the corrugations, in order to minimize its effect on
the optical coupling and to maximize the thermal efficiency of the
device (see Figure \ref{fig:absgeom}).

\noindent Optimal coupling efficiency requires careful impedance
matching of the PSB to the coupling structure. For a given
absorber geometry, impedance matching is achieved by controlling
the thickness of the absorber metallization and therefore the cold
surface impedance of the device.  The surface impedance resulting
in the highest power coupling efficiency is dependent not only on
the geometry of the coupling structure and the frequency of
operation, but also on the geometry of the absorber.

\noindent This fact can be understood in terms of classical
transmission line theory.  For an idealized transmission line of
characteristic impedance $Z_0$ terminated in a load with impedance
$Z_{\mathrm{load}}$ one expects a reflection amplitude ($S_{11}$,
normalized to input) of

\[
| ~ S_{11}| = \left| \frac{Z_0 - Z_{\mathrm{load}}}{Z_0 +
Z_{\mathrm{load}}} \right|
\]

\noindent In the limit that the characteristic spacing of the
absorber grid is much less than a wavelength, the transmission
line model is applicable to the PSB coupling.  In order to
parameterize the power coupling efficiency as a function of
bolometer surface impedance, we use the functional form suggested
by transmission line theory,
\begin{equation}
\eta \left(Z_{\mathrm{abs}}\right) = f \times \left[ 1 - \left(
\frac{Z_0 - Z_{\mathrm{abs}}}{Z_0 + Z_{\mathrm{abs}}}\right)^2
\right]
 \label{pwrcouple}
\end{equation}
where $f$ and $Z_0$ are free parameters, while $Z_{\mathrm{abs}}$
is determined by the processing of the absorber.

\begin{figure}[tbp]
\begin{center}
\rotatebox{90}{\scalebox{0.35}{\includegraphics{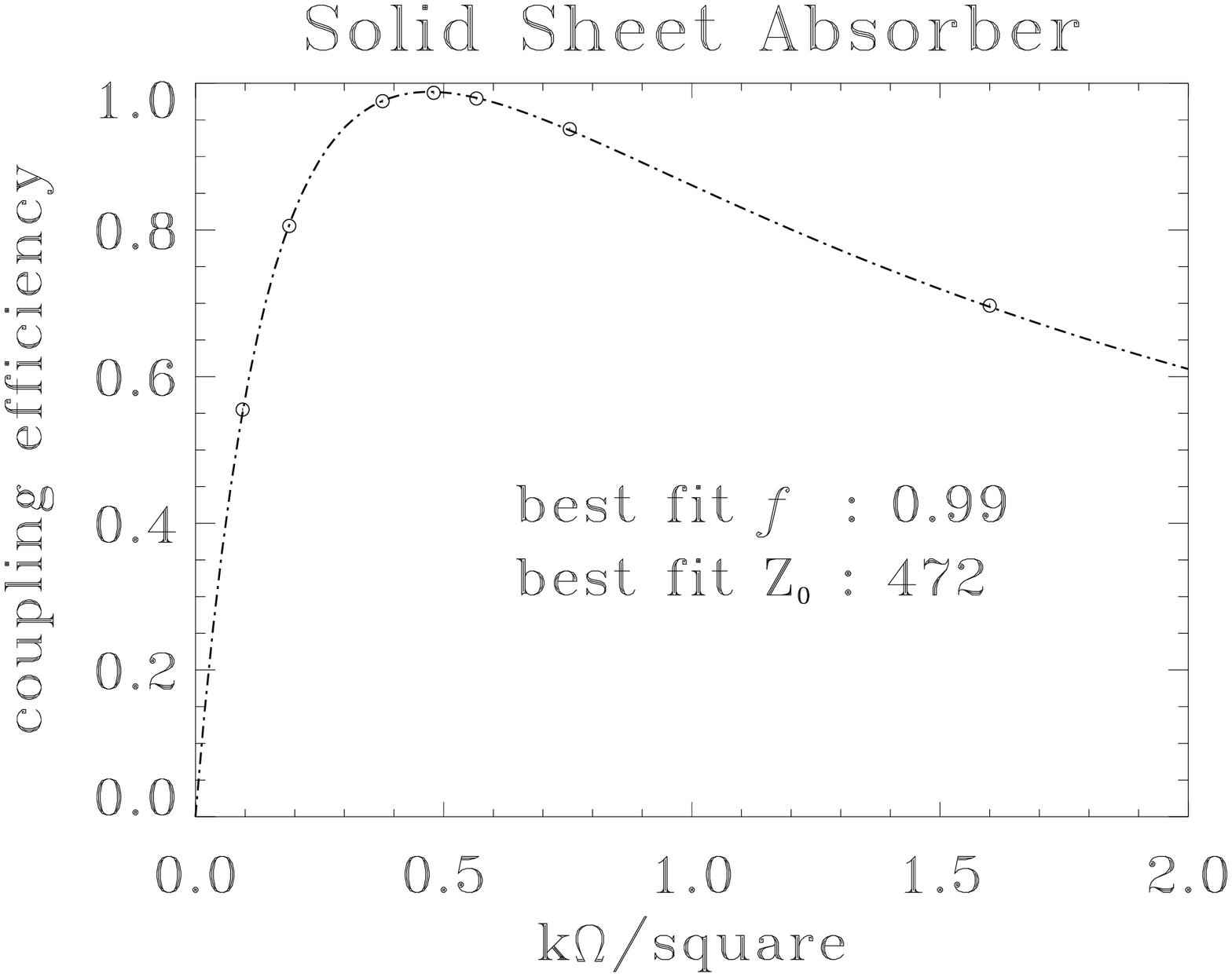}}}
\rotatebox{90}{\scalebox{0.35}{\includegraphics{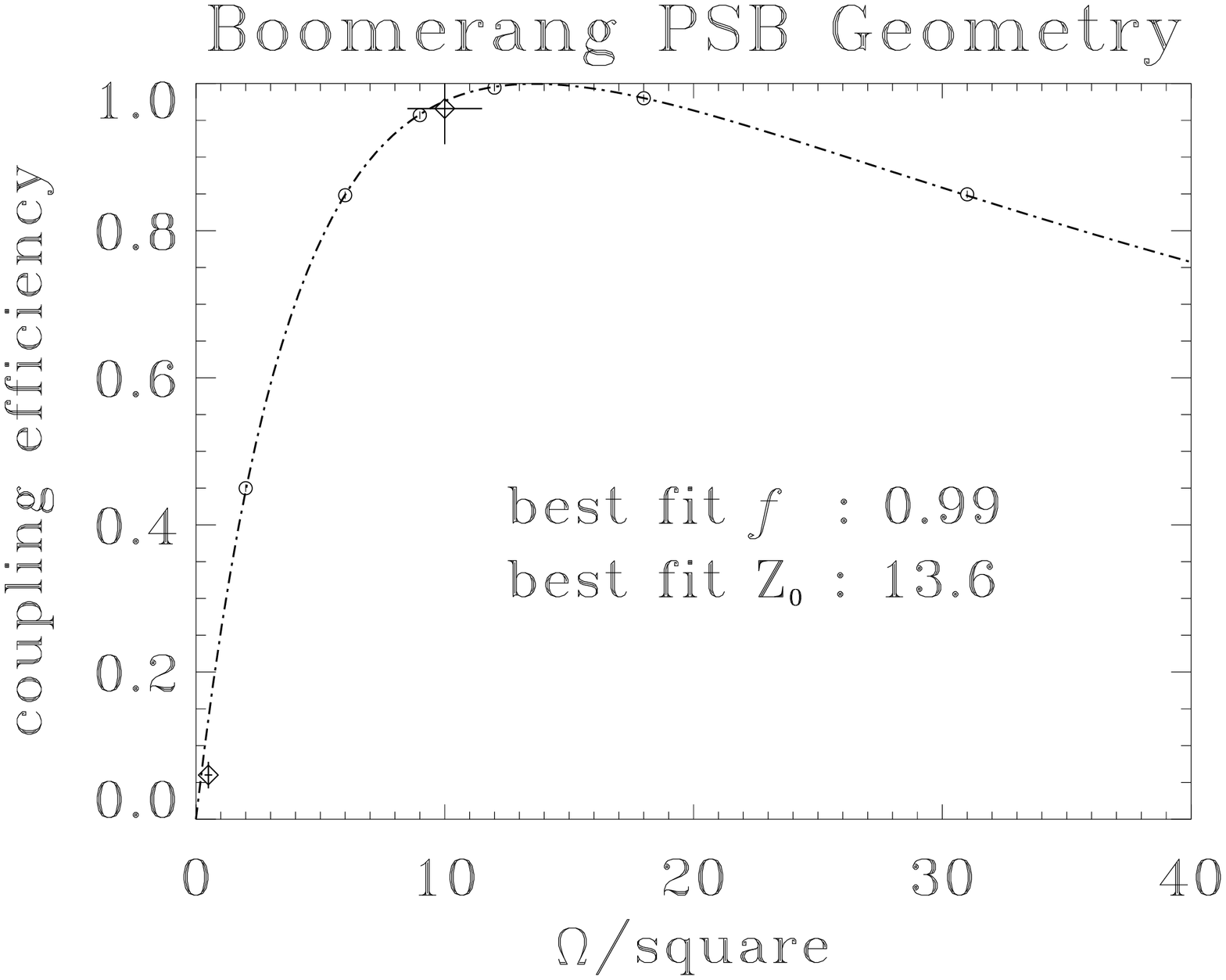}}}
\caption{At left, the numerically determined coupling efficiency,
$\eta (Z_\mathrm{abs})$, of a solid sheet of uniform surface
impedance mounted in a PSB coupling structure. The open circles
are the numerical results. A function of the form of Equation
\ref{pwrcouple} is fit to the points to determine the peak
coupling efficiency and characteristic impedance of the structure
at the center frequency of the operational band.  At right, the
same numerical procedure is performed for the particular geometry
of the \boom PSBs.  Note the change in scale of the abscissa, from
k$\Omega$ to $\Omega$, which can be understood in terms of the
filling factor of the PSB absorber (see Equation
\ref{eqn:sheetabs}). For the \boom geometry, two absorber
metallizations were fabricated and tested, corresponding to
approximately 0.5 $\Omega$/square and 10 $\Omega$/square.  The
measured coupling efficiency of the two types of devices are
indicated by the open diamonds, and are in excellent agreement
with the numerical result. The error bars represent the
uncertainty in the cold impedance of the metallization layer and
the transmission of the window material and optical filters.}
\label{fig:sheetabs}
\end{center}
\end{figure}

\noindent In order to determine the optimal absorber impedance for
our devices, numerical computations of
$\eta\left(Z_{\mathrm{abs}}\right)$ were made for discrete values
of $Z_{\mathrm{abs}}$ using the Ansoft HFSS package.\cite{ansoft}
These results were used to determine $f$ and $Z_0$ for each of six
device geometries at the center frequency of our band. The
geometries included a solid sheet, a \lq\lq spider-web\rq\rq
bolometer, and four types of PSBs, all of which were found to be
well fit by the functional form of Equation \ref{pwrcouple}. The
results of the numerical procedure for two of the devices, a solid
sheet and the PSB design chosen for \boomn, are shown at the left
and right of Figure \ref{fig:sheetabs}, respectively. The best fit
surface impedances ranged from $470~\Omega/$square for the solid
sheet to nearly $5~\Omega/$square for the device with the finest
absorber leg widths.

\noindent In order to validate both the numerical procedure and
the application of transmission line theory to the PSB coupling,
optical efficiency measurements of devices with physical surface
impedances of $\sim 0.5 ~\Omega$/square and $\sim 10
~\Omega$/square were made.  The numerical, theoretical, and
measured efficiencies are compared in Figure \ref{fig:sheetabs} at
right.  The well matched devices showed optical efficiencies
nearly a factor of four higher than the low impedance devices, and
validate the applicability of transmission line theory to the PSB
coupling structure.

\noindent Similar numerical simulations were made of a smooth
walled conical feed exiting into an integrating cavity. This
structure was found to deviate significantly from the transmission
line theory and showed little dependence on the details of the
absorber geometry, presumably because the abrupt discontinuity in
the boundary conditions at the entrance to the cavity dominate the
return loss.

\noindent The results of the numerical computations suggest that
it is not possible, in principal, to analytically calculate the
optimal absorber surface impedance for an arbitrary geometry due
to the fact that the characteristic impedance of the structure,
$Z_0$, is itself dependent on the web geometry. However, in the
limit that the absorber geometry approaches a solid resistive
sheet,\, the characteristic impedance of the PSB coupling
structure approaches $Z_0 \rightarrow 470$ Ohms. This numerically
determined characteristic impedance is greater than that of free
space (377 Ohms), as is to be expected for an electrically small
structure.

\noindent An approximate relationship for the optimal target
surface impedance of a given absorber geometry can then be
obtained in the following way. Imagine making $N$ infinitesimally
thin slices through a solid sheet absorber of this surface
impedance. Clearly $Z_0$, the optimal surface impedance of a solid
sheet, is still the appropriate target impedance. If one then
reduces the width, $w$, of each of the $N+1$ legs of the absorber
without adjusting the physical surface impedance from $Z_0$, the
{\em optically} active area will no longer have the same
characteristic impedance.  For a given absorber diameter $D$ and
center to center leg spacing $g = \frac{D}{N+1}$, each leg will be
acting for an optical area which is fixed ($g\times L$), while its
geometric area ($w\times L$) is clearly not. Therefore, one would
expect that the optimal target surface impedance would scale as
the ratio of geometric area to optical area,
\begin{equation}
 Z_{\mathrm{abs}} \approx Z_0 \frac{w}{g} = Z_0 \frac{w(N+1)}{D}
\label{eqn:sheetabs}
\end{equation}
so that in the limit that the filling factor approaches unity the
surface impedance, $Z_\mathrm{abs}$, approaches $Z_0$.  For $g$\rq
s less than $\lambda/5$, we have found good agreement between this
approximate relation and numerical calculations of the optimal
surface impedance of a given absorber geometry.  For reference,
the \boom devices have 23 absorber legs of width $w \simeq 3
\mu$m, and a 2.6 mm absorber diameter (see Table \ref{tbl:bolos}).
Using the $Z_0$ determined from a solid sheet absorber,\,
application of Equation \ref{eqn:sheetabs} yields an optimal
impedance of $12.5~\Omega/$square, versus the numerically
determined value of 13.6.

\begin{figure}[tbp]
\begin{center}
\rotatebox{90}{\scalebox{0.35}{\includegraphics{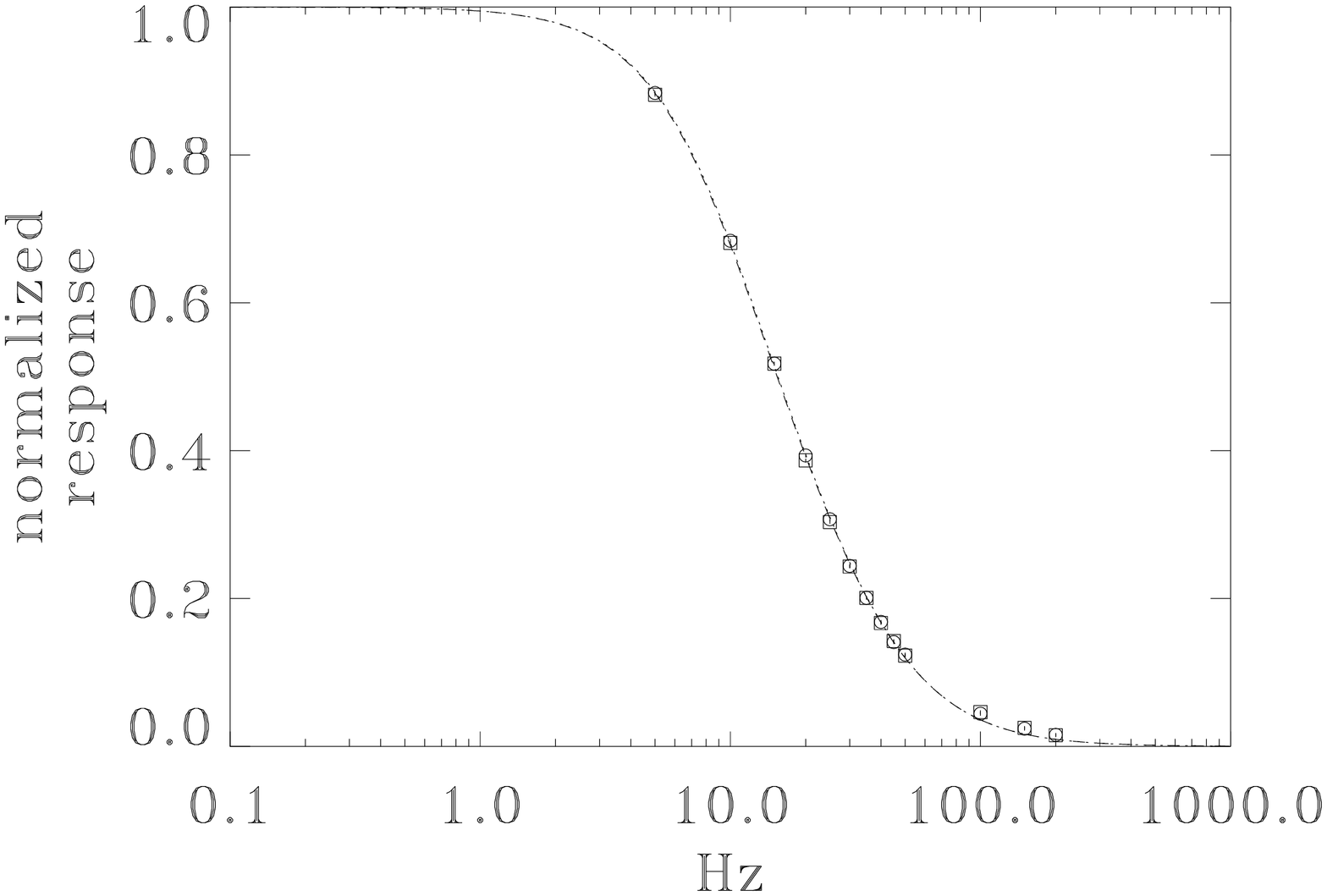}}}
\rotatebox{90}{\scalebox{0.35}{\includegraphics{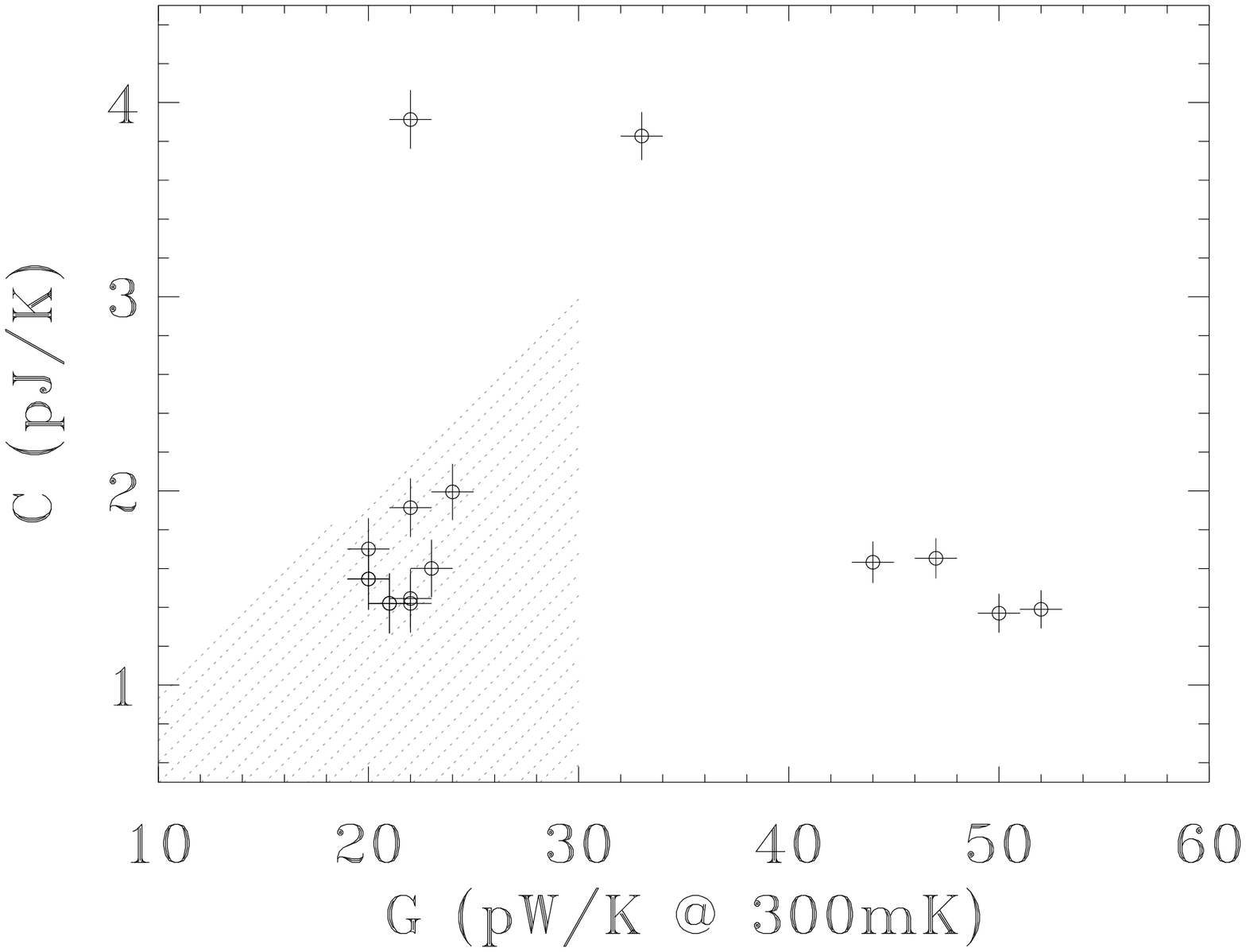}}}
\caption{The thermal properties of the prototype PSBs. At left,
the measured response of a PSB pair to a chopped optical source.
The two devices, which are over-plotted, are very well matched in
$\tau$ and therefore well matched in both heat capacity and
thermal conductivity.  At right, a plot of the measured heat
capacity versus thermal conductivity for a variety of PSB devices
at 300mK. The cross-hatched region is the region of parameter
space acceptable for \boomn\rq s 150 GHz PSBs.}
\label{fig:heatcap}
\end{center}
\end{figure}

\subsection{Thermal Design}

The bolometer design must ensure that the power deposited on the
absorber is efficiently detected by the thermistor.  The parasitic
thermal conductivity of the \sini supports and inefficient heat
transport across the absorber can potentially result in a loss of
detector efficiency if a significant fraction of the optical power
dissipated in the absorber does not result in a temperature rise
of the thermistor.  For this reason, several features were built
into the PSB design to ensure sufficient thermal efficiency.

\noindent The electrical leads constitute the dominant thermal
path from the bolometer to the temperature bath. The thermistor is
located on a pad which is heat sunk to both the leads and the
absorber via a thermally conductive ring surrounding the absorber.
For the extremely low background bolometers with thermal
conductivities below 30 pW/K, the \sini beam supporting the leads
has been found to be a significant contribution to the total
thermal conductivity.  The scaling of the thermal conductance with
temperature for these devices is consistent with a $T^3$
dependence, as described in Holmes, et al\cite{holmes98}.

\noindent A finite element thermal model of the bolometer design
was implemented in order to analyze the thermal efficiency of the
bolometer design. The thermal conductance and heat capacity of the
metallization and \sini structures used by the model were obtained
empirically in independent analysis of similar
structures.\cite{holmes98,pdm97} The thermal model allows the
calculation of the temperature rise at the thermistor resulting
from power (DC and AC) dissipated on the absorber, the bolometer
response to a transient pulse of power deposited on the absorber
(similar to a cosmic ray hit), and the ratio of power flow through
the thermistor to the total power deposited on the bolometer.

\noindent The response of the device to a chopped source was
measured over a $\sim 1$~kHz bandwidth.  The measured bolometer
response agrees very well with the finite element model.  The
calculated ratio of rise time to relaxation time for a transient
pulse is $\sim 10 \%$ at 400 mK, which indicates that the crossing
time for temperature fluctuations across the absorber is much less
than the thermal time constant of the bolometer. Finally, the
results of the thermal analysis indicate that the ratio of power
flow through the thermistor to the total power deposited on the
mesh is $0.97$ for the prototype devices.

\begin{figure}[tbp]
\centering
\rotatebox{90}{\scalebox{0.34}{\includegraphics{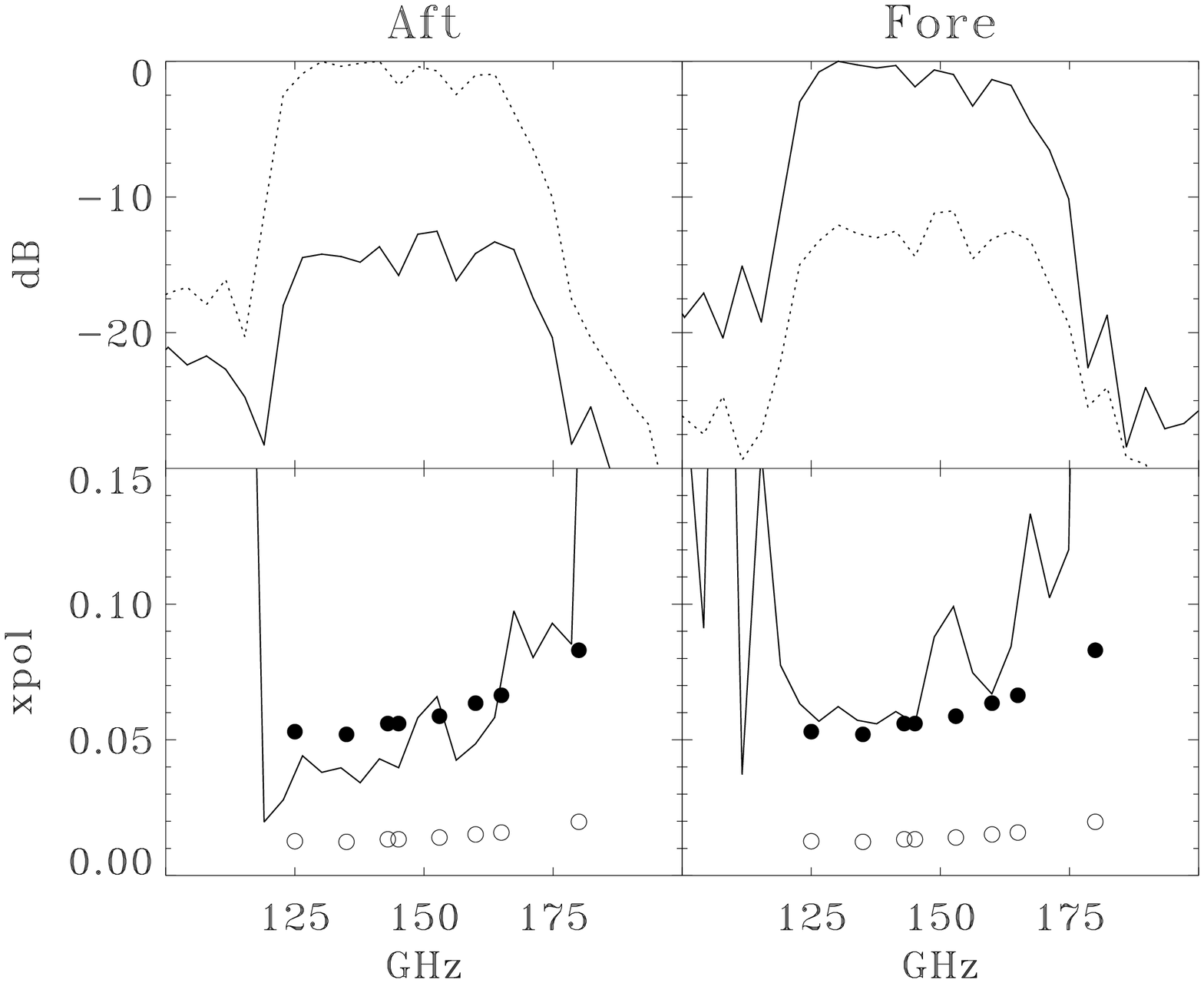}}}
\hspace{5mm}\rotatebox{90}{\scalebox{0.36}{\includegraphics{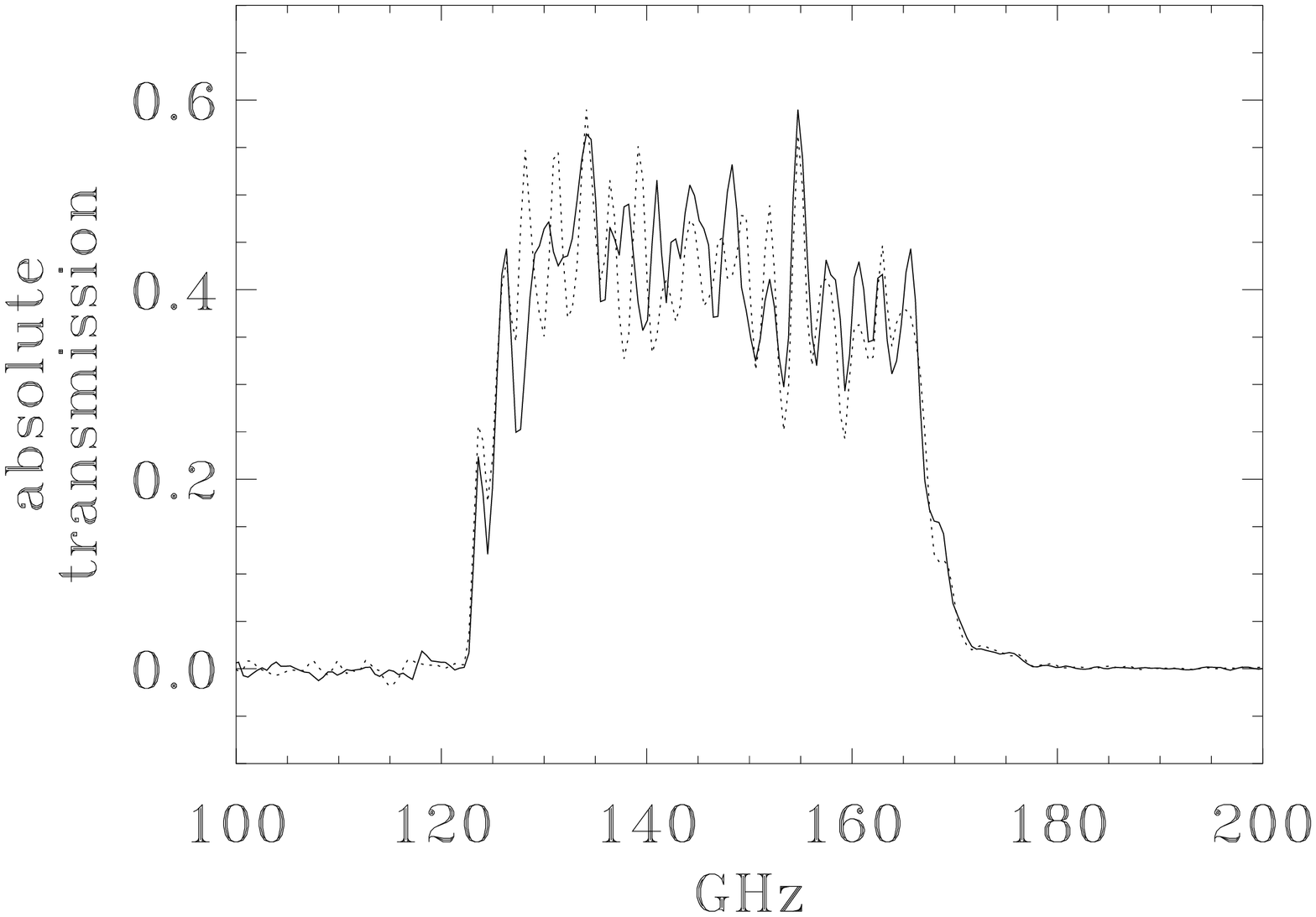}}}
\caption{At left, the measured cross-polar performance of a
prototype (Type 04) PSB in the \boom testbed. The upper panels
show the measured spectra in the two orthogonal directions defined
by the PSB pair. Below, the measured cross-polar leakage,
$\epsilon_i$, as a function of frequency for a PSB pair. The
spectral resolution of the measurement is 4 GHz and the signal to
noise is only significant within the passband, from 124 to 168
GHz. The filled circles indicate the single frequency results of
the HFSS simulations. The open circles are the simulation results
for the device geometry designed for \planckn. At right, the
measured optical efficiency of the \boom PSBs.  The spectra are
taken with a Fourier transform spectrometer at 500 MHz resolution.
The spectral normalization is obtained by measuring the electrical
power difference at constant resistance with 77K, 90K, 273K, and
300K beam filling loads.  The spectra for the two bolometers in a
PSB pair are shown, one with a solid line, the other a dotted
line.} \label{fig:xpol_bands}
\end{figure}

\section{Polarized signal analysis}
\label{sec:readout}

 The two bolometers in a PSB pair can be
read out either in a bridge circuit or individually.  A bridge
circuit has the advantage of measuring the difference signal
directly through a single amplifier chain, however it requires
close matching of the thermistor properties and does not provide a
precise simultaneous measurement of $I$. For the study of the
cosmic background radiation, it is highly advantageous to measure
both the polarized and unpolarized components
simultaneously.\cite{oc02} In the remainder of this paper we will
consider the signals received by the bolometers individually. An
analysis of the individual readout illustrates the fundamental
properties of the PSBs and how they can be interpreted in terms of
the Stokes parameters.

\subsection{PSB readout}

\noindent In a suitably defined coordinate system the measured
signal voltage is
\begin{equation}
v_i = S_i \cdot\frac{1}{2}~ \int d\nu ~\lambda^2 ~\eta_i ~F_i(\nu)
~\left[ (1+\epsilon_i)\cdot I ~ \pm ~ (1-\epsilon_i) \cdot (Q\cos
(2\alpha_i) + U\sin(2\alpha_i)) \right] \label{eqn:sigs}
\end{equation}
where the sign of the polarized term is associated with the
copolar $(+)$ or crosspolar $(-)$ device, $S_i$ is the voltage
responsivity of the bolometer, $\lambda^2$ is the throughput
($A\Omega$) of the system, and the product $\eta_i ~ F_i(\nu)$ is
the absolute spectral transmission. The crosspolar leakage,
$\epsilon_i$, is the quantity measured in Figure
\ref{fig:xpol_bands}, and can be thought of as the response of a
device to a 100\% linearly polarized source oriented perpendicular
to the design axis of sensitivity. Equivalently, one could define
a polarization efficiency of the detector, $\xi \equiv
1-\epsilon$, which, as is shown below, enters as a multiplicative
factor in the overall efficiency of the system to polarized
radiation. The alignment angle, $\alpha_i$, is the orientation of
the bolometer in the coordinate system which defines $Q$ and
$U$.\footnote{Kaplan and Delabrouille have investigated the impact
of uncertainty in these paramters via Monte Carlo estimation.  For
reference, it is relatively easy to measure both $\epsilon$ and
$\alpha$ to a few percent precision, which roughly correspond to
the highest precision considered in that
analysis.\cite{jd01,jk01}} The bolometer voltage responsivities,
$S_i$, are dependent on the background power, the temperature of
the bath, the thermal conductance, the spectral bandpass, the
optical efficiency, and the thermistor properties.

\noindent Ideally, the two bolometers would have $\epsilon=0$ and
would be oriented exactly $90^\circ$ with respect to one another,
however in practice the two devices exhibit crosspolar response at
the few percent level and are typically orthogonal to within
$1.5^\circ$. It is important to note that a single PSB cannot
fully characterize the polarization of the signal without some
method of rotating $Q$ into $U$, or vice versa.  Ideally one would
modulate the polarized component of the radiation, as with a
wave-plate or Faraday rotator, in order to unambiguously measure
both the Stokes $Q$ and $U$ parameters from a single feed. In
practice, instruments such as \boom and \planck will rely on
scanning through several orientations on the sky in order to build
up information on both Q and U.

\noindent In a suitably defined coordinate system, $\alpha_1
\simeq 0^\circ$ and $\alpha_2 \simeq 90^\circ$.  Therefore, so
long as $\epsilon_i \ll 1$, a measurement of $I$ is obtained by
summing the two signals while a measurement of $Q$ is obtained by
taking the difference. For simplicity, we approximate Equation
\ref{eqn:sigs} by removing $I, Q$, and $U$ from the frequency
integral and introduce the factor $\mathcal{F}_i \equiv \int d\nu
~\lambda^2 ~\eta_i ~F_i(\nu)$. Taking the signal difference
yields,
\begin{equation}
(v_1-v_2) = \gamma\cdot I + \delta\cdot Q \label{eqn:pdiff}
\end{equation}
where we have made the definitions,
\begin{equation}
 \gamma \equiv \left[ S_1 \cdot \mathcal{F}_1 \cdot (1+\epsilon_1)~-~S_2 \cdot
\mathcal{F}_2 \cdot (1+\epsilon_2)\right] \label{eqn:gamma}
\end{equation}
\begin{equation}
 \delta \equiv \left[ S_1 \cdot \mathcal{F}_1 \cdot (1-\epsilon_1)~+~S_2 \cdot
\mathcal{F}_2 \cdot (1-\epsilon_2)\right]
\end{equation}
The parameter $\epsilon$ is identical to a degradation in
sensitivity to polarization.  The common mode rejection ratio
(CMRR) of the PSB receiver depends crucially on the stability of
the coefficient $\gamma$ in Equation \ref{eqn:gamma}.  For studies
of the CMB, the dominant unpolarized signal, $I$, will be that of
the temperature anisotropy. For the currently favored $\Lambda$CDM
cosmology, the ratio of the polarized to unpolarized power is
expected to range from 2-6\% over the angular scales of interest,
which determines the level of stability required of the parameter
$\gamma$.\footnote{Specifically, $\langle EE\rangle / \langle
TT\rangle$, the ratio of gradient mode polarization to temperature
anisotropy bandpowers over angular scales of  $\sim 2^\circ$ to
$10'$, or $\ell \simeq 400-2000$.}

\subsection{Polarization efficiency}

\noindent From Equation \ref{eqn:gamma}, it is evident that the
stability of $\gamma$ depends on the stability of both the cross
polar leakage and the calibration. The intrinsic polarization
efficiency of a PSB is a property of the geometry of the absorber
and coupling structure, the metallization of the absorber, and the
spectral bandpass.\footnote{Of course, the efficiency of a given
instrument depends as well on the properties of optical
arrangement, including the filters, feed elements, reflectors, and
window material, etc. At the percent level, we have observed a
degradation of polarization efficiency with increased thickness of
the window material.} The numerical simulations provide some
intuition about the origin of this cross polar response.  For a
single frequency, one can integrate the component of the Poynting
vector normal to two separate surfaces, each of which fully
enclose one of the bolometers.  The model used includes the
detailed geometry of the absorber, the thermistor, the electrical
leads, and all support structures. For a given source
polarization, a comparison of the total integrated flux through
these surfaces yields the ratio of power absorbed on each
bolometer.  The results of this numerical procedure for five PSB
absorber geometries are presented in Table \ref{tbl:bolos}.  In
all cases, the surface impedance which provided the maximum
polarization efficiency was found to coincide with value yielding
the peak power absorption.  A deviation from optimal surface
impedance of~$\sim 20\%$~was found to result in a degradation of
the polarization leakage by~$~\sim 12\%$.

\noindent Measurements of the cross polarization are in good
agreement with the numerically determined values.  Two methods
were used to measure the cross polar response of the system. A
polarized Fourier Transform Spectrometer was used to study the
spectral dependence of the polarization efficiency. The left panel
of Figure \ref{fig:xpol_bands} shows the measured spectrum of the
cross polar response for a prototype PSB installed in the \boom
testbed at 4 GHz resolution. The filled circles represent single
frequency results from numerically integrating the Poynting flux.
A second test used a chopped thermal (Raleigh-Jeans) source and
polarizing grid to measure the broadband response as a function of
grid orientation. The latter measurement will generally result in
a lower polarization efficiency than will result in a measurement
of the CMB due to the relatively flat spectrum of the cosmological
signal and the rising spectrum of the polarization leakage.

\noindent As is evident from Figure \ref{fig:pflux}, a significant
fraction of the cross polar response of the prototype devices
originates from power dissipation at the edge of the mesh. Another
factor found to influence the cross polar response is the line
spacing, $g$. The calculated cross polar level generally declines
with a more densely packed grid.  In order to determine the
effect, if any, of the \sini mechanical supports running
perpendicular to the absorbing legs on the cross polarization, the
support legs of a prototype device were removed by laser oblation
after optical testing.  Retesting the device showed no change in
the level of polarization leakage. The polarization leakage,
$\epsilon$, is a material property of the device and is entirely
stable.

\begin{table}[t]
\caption{Bolometer design parameters for the four prototype device
geometries (Types 01--~04), and the \planck 143P geometry (type
05). The values of $Z_0$ and $\epsilon$ quoted in the last two
columns represent the numerically determined values at 150 GHz.
~$^\S$ The polarization leakage, defined by Equation
\ref{eqn:sigs} is calculated at optimal absorber impedance.
$^\dagger$ The devices geometries used in \boomn. $^\ddagger$ The
device designed for \planckhfin\rq s 143P channels.}
\label{tbl:bolos}
\begin{center}
\begin{tabular}{l|c|c|c|c|c} %% this creates two columns
%% |l|l| to left justify each column entry
%% |c|c| to center each column entry
%% use of \rule[]{}{} below opens up each row
%\hline
\rule[-1ex]{0pt}{3.5ex}  PSB & mesh diameter  & line width  & line spacing & char. imped. & xpol. leakage$^\S$\\
\rule[-1ex]{0pt}{3.5ex}  Geometry  & $d$~[~mm~] & $w~[~\mu$m~] & $g~[~\mu$m~]& $Z_0$~[~$\Omega$/square] & $\epsilon$~[~\%~] \\
\hline\hline
\rule[-1ex]{0pt}{3.5ex}  Type $01^{\dagger}$  & 2.6 & 3 & 108 & 13.6 & 2.6 \\
\hline
\rule[-1ex]{0pt}{3.5ex}  Type $02^{\dagger}$  & 2.6 & 5 & 108 & 22.5 & 2.3 \\
\hline
\rule[-1ex]{0pt}{3.5ex}  Type 03  & 2.6 & 3 & 325 & 4.9 & 5.6 \\
\hline
\rule[-1ex]{0pt}{3.5ex}  Type 04  & 2.6 & 3 & 217 & 6.8 & 5.4 \\
\hline
\rule[-1ex]{0pt}{3.5ex}  Type $05^{\ddagger}$  & 3.2 & 3 & 100 & 15.4 & 1.3 \\
%\hline
\end{tabular}
\end{center}
\end{table}
%\newpage

\noindent Finally, an additional effect which can be important is
the role of electrical crosstalk between readout channels.
Electrical shielding of the bias and signal lines is required to
minimize this effect, especially between the two devices within a
pair. This crosstalk is indistinguishable from cross polar
leakage, $\epsilon$, and must be both carefully avoided and
characterized.

\subsection{Relative calibration}
We have shown that the stability of the relative calibration of
the two polarization senses is critical in order to achieve an
acceptably small systematic contribution to the signal. This
situation is by no means unique to a PSB receiver. Coherent
receivers are susceptible to large gain fluctuations and phase
error in the front end amplifiers which are analogous, though
typically far larger in amplitude, to fluctuations in the product
$S_i \cdot \mathcal{F}_i$. For this reason most correlation
receivers used in measurements of the CMB use some sort of
synchronous demodulation of the signal at frequencies above that
of the gain fluctuations. As we will see, the PSB architecture is
designed to realize stability in the relative calibration at
frequencies as low as a few tens of milliHertz.

\noindent Fluctuations in the relative calibration arise from
mismatched effective throughput, $\mathcal{F}_i$, as well as
drifts in the voltage responsivity,~$S_i$. The right panel of
Figure \ref{fig:xpol_bands} shows band averaged optical
efficiencies, esssentially $d\mathcal{F}_i/d\nu$, of the \boom
PSBs. The efficiencies are matched to within a few percent, while
the band edges are matched to better than 0.2\%. The edge of the
band is determined by the waveguide cutoff of the feed, while the
upper edge is defined by metal mesh resonant filters.\cite{lee96}
Due to the fact that the optical path for each polarization sense
is identical, the voltage responsivity will dominate variations in
the calibration.  As we will see, the great advantage of the PSB
architecture is the fact that the factors which determine the
relative responsivity are intrinsic to the physical arrangement,
and are therefore extremely stable.

\noindent Any number of authors have discussed the properties of
bolometric receivers; the authoritative treatment is given in the
series of papers by Mather.\cite{mather82,mather84}  It can be
shown that the voltage responsivity of a bolometer under a given
optical load is completely determined by five quantities, the
spectral bandpass, the quantum efficiency, the temperature of the
bath, the thermal conductance to the bath, and the properties of
the thermistor material. We have discussed the first two items,
and have shown them to be well matched and extremely stable. The
physical proximity of the devices, separated by $< 60\mu$m on a
beryllium copper housing, ensures that the two devices operate
from a common bath temperature.  The thermal conductance to the
bath is determined by the uniformity of the metal deposition of
the electrical leads, and is matched to within $\sim 10\%$ for the
devices fabricated on a single wafer for the \boom PSBs. Better
matching is possible for devices with higher thermal conductivity
than those designed for \boomn, $G \simeq 20$ pW/K at 300 mK.  The
final contribution to the difference in responsivity of devices
within a PSB pair is the variation in the thermistor properties
between the two devices. The NTD germanium thermistors used in
this work have a resistance versus temperature well described by
$$ R(T)=R_0 e^{\sqrt{\frac{\Delta}{T}}} $$
Variation in the parameter $\Delta$, which is a property of the
doping of the germanium, is found to be much smaller than the
variation in $R_0$, which is dependent on the geometry of the
chip. However, both parameters are intrinsic to the chip, and
likewise are entirely stable.

\section{The prototype receiver}

\begin{figure}[tbp]
\centering
%\rotatebox{0}{\scalebox{0.2}{\includegraphics{psb_noise.prn}}}
\rotatebox{0}{\scalebox{0.8}{\includegraphics{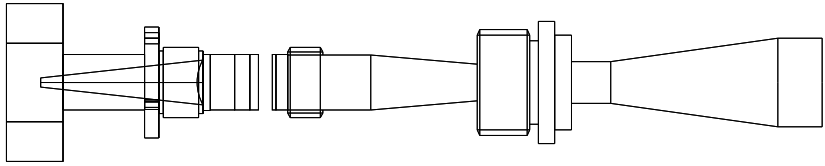}}}
\caption{The \boom 150 GHz PSB feed assembly.  The 2K feed, though
not a necessary component of the PSB design, is a convenient
method of including additional filtering and control of the beam
without loss of performance.  Both the 2K and sub-K feeds are
fully corrugated and profiled to provide high fidelity
transmission of the polarized signal, symmetric E and H plane
beams, and relatively high gain in a compact system.}
\label{fig:proto}
\end{figure}

The experimental results reported in this work are primarily the
results of the development program for the new \boom focal plane.
The \boom receiver is a prototype of the \planckhfi 143 GHz
polarized receiver. We include for completeness the details of
this system, so that the data may be interpreted in the proper
context.

\noindent In addition to the feed element described in Section
\ref{sec:optics}, the prototype optical system consists of a
profiled, corrugated back to back feed which is cooled to 2K. This
feed defines the lower edge of the operational band, and couples
to the reflector system.  The high pass edge of the sub-K assembly
is designed to be $\sim 10$\% below that of the 2K feed in order
to assure that both the throughput and beam definition are
provided by the 2K feed element. Two different 2K feeds were
designed and tested, each of which is corrugated in both the front
and back sections.  Neither of these feeds were found to
contribute to the cross polarization or to degrade the optical
efficiency when properly configured.  However, replacing the sub-K
feed with a smooth-walled but otherwise identical feed resulted in
a two-fold increase (roughly 5\%, absolute) in the crosspolar
response, $\epsilon$, of a PSB receiver tested in the \boom focal
plane.

\noindent The system has achieved high end-to-end efficiency. A
band average efficiency of nearly 40\% is measured by calculating
the power difference observed between beam-filling loads which are
controlled at temperatures of 273 and 300K, as well as 77K and
90K. The right panel of Figure \ref{fig:xpol_bands} shows the
normalized transmission of the system as measured in the \boom
system. This efficiency is not corrected for any loss in the
vacuum window material, the blocking filters at 80K and 2K, or the
sub-K or 2K feeds.

\section{Conclusion}

We have demonstrated a 300 mK bolometric receiver which is
intrinsically sensitive to linear polarization over a 33\%
bandwidth.  The general design is scalable from $\sim 60-600$ GHz.
This design benefits from reduced susceptibility to systematic
effects due to the common filtering, matched beams on the sky,
matched time constants, stable relative responsivities, and
matched end-to-end efficiencies of each sense of linear
polarization.   Unlike coherent correlation polarimeters, this
receiver simultaneously measures the polarized and unpolarized
components of the signal with comparable sensitivity. The design
minimizes the size and weight of the receiver, making it
especially appropriate for orbital and sub-orbital compact
feedhorn arrays. A general method of reliably calculating the
optimal absorber impedance for a bolometric detector is presented.
The measured performance of the system is in good agreement with
the results of the numerical modelling.

\newpage
% http://www.astro.caltech.edu/~wcj/anim_psb.html
\acknowledgments     %>>>> equivalent to \section*{ACKNOWLEDGMENTS}
\noindent The authors would like to acknowledge Peter Ade and
Carole Tucker, who kindly provided the optical filters for the
development program. Tom Montroy and Ted Kisner provided
invaluable information about the performance of the PSBs installed
in the \boom focal plane, and Eric Torbet has measured the
polarized spectra of the PSBs integrated in the \boom focal plane.
WCJ would like to thank Kathy Deniston for facilitating this
development effort, Goutam Chattopadhyay for his useful comments
on the use of the HFSS software package, and Jonas Zmuidzinas,
Marcus Runyan, and Brian Jones for helpful discussions. Thanks to
Paolo deBernardis for making the authors aware of the early work
of Caderni, et al. RSB is currently with the European Space
Agency, ESTEC, Noordwijk, The Netherlands. William Jones is
supported through NASA GSRP fellowship NGT5-50278. We are grateful
for the generous support of this effort by The Caltech Discovery
Fund.

%%%%% References %%%%%

\end{document}